\documentstyle[PASJadd,epsf]{PASJ95}
\draft

\large\normalsize

\begin{document}
\title{X-Ray Evidence of an AGN in M82}

\author{Hironori {\sc Matsumoto}\\
{\it The Institute of Physical and Chemical Research (RIKEN),} \\
{\it 2-1 Hirosawa, Wako, Saitama 351-0198}\\
{\it E-mail(HM): matumoto@crgotch.riken.go.jp}\\
and\\
Takeshi Go {\sc Tsuru}\\
{\it Department of Physics, Faculty of Science, Kyoto University,}\\
{\it Sakyo-ku, Kyoto 606-8502} \\
{\it E-mail(TGT): tsuru@cr.scphys.kyoto-u.ac.jp}\\
}

\abst{
An X-ray spectrum of the famous starburst galaxy M82
consists of three components: soft, medium, and hard
components (Tsuru et al. 1997). The spectrum of the hard
component, which is spatially unresolved, is well
represented by an absorbed thermal bremsstrahlung, or an
absorbed power-law model. However the origin of the hard
component was unclear.  Thus, we made a monitoring
observation with ASCA in 1996. Although the X-ray flux of
the soft and medium components remained constant, a
significant time variability of the hard component was found
between $3\times10^{40}$ erg/s and $1\times10^{41}$ erg/s at
various time scales from 10 ks to a month. The temperature
or photon index of the hard component also changed. We
proved that the spatial position of the hard component is
the center of M82.  The spectrum of the variable source
obtained by subtracting the spectrum of the lowest state
from the highest state suggests the strong absorption of
$N_{\rm H} \sim 10^{22}$ cm$^2$, which means the variable
source is embedded in the center of M82. All these suggest
that a low-luminosity AGN exists in M82. }

\kword{Galaxies: active --- Galaxies: individual (M82) 
--- Galaxies: X-rays  --- X-rays: spectra}

\maketitle
\thispagestyle{headings}

% section 1
\section{Introduction}

Because M82 is thought to be an archetypical starburst
galaxy, many X-ray observations of M82 have been made. The
ASCA first observation of M82 was conducted in 1993.  Tsuru
et al. (1997) analyzed the ASCA spectrum and found that it
consists of three components: soft, medium, and hard
components.  The soft and medium components showing emission
lines from various elements are thermal origin at
temperatures of $\sim$ 0.3 and $\sim$ 1 keV. The ASCA images
of the soft and medium components are extended compared with
the ASCA point spread function (PSF). Thus their origin
would be the galactic wind driven by stellar winds from
massive stars and supernovae.

The hard component can be well described by either a
power-law model (the photon index is $\Gamma \sim$ 1.7) or a
thermal-plasma model (the temperature is $kT \sim$ 14 keV).
This component is dominant in the X-ray spectra above the 2
keV band, and its X-ray luminosity in the 2 -- 10 keV band
is $\sim\ 3\times10^{40}$ erg/s.  Tsuru et al. (1997)
compared the ASCA flux in the 2 -- 10 keV band with Ginga
(Tsuru 1992) and EXOSAT (Schaaf et al. 1989), and found time
variability. Furthermore, the spatial extent of the hard
component is consistent with a point source within the ASCA
angular resolution. All these may suggest that the origin of
the hard component is a low-luminosity AGN of M82. However,
since Ginga and EXOSAT are non-imaging detectors, possible
contamination from other hard sources is not excluded.
Therefore, it is unclear whether the flux change can be
attributed to the hard component or not. Since the peak
position of the hard component does not agree with the soft
component, it is unclear whether the peak of the hard
component agrees with the center of M82 where a luminous
X-ray point source showing time variability exists (Watson
et al. 1984; Collura et al. 1994; Bregman et al. 1995;
Strickland et al. 1997). Furthermore, the Ginga spectrum can
be fitted with a thermal model but cannot be fitted with a
power-law model (Tsuru 1992), which is different from
typical X-ray spectra of AGNs. The same conclusion was
suggested by Cappi et al. (1998) using BeppoSAX data.
Thus the origin of the hard component is still debatable.

The first key to reveal the origin of the hard component is
to clarify whether it shows time variability or not. The
second key is to detect the iron K-line emission and to
determine its central energy. The central energy of the line
can be direct evidence. For these purposes, we made a
monitoring observation of M82 with ASCA in 1996. The total
exposure time including the observation in 1993 is about
$1.5\times10^5$ s.

Throughout this paper, a distance of 3.25 Mpc to M82 is assumed.  The
number of atoms per hydrogen for the cosmic metal abundance adopted in
this paper are $9.77\times10^{-2}$ for He, $3.63\times10^{-4}$ for C,
$1.12\times10^{-4}$ for N, $8.51\times10^{-4}$ for O,
$1.23\times10^{-4}$ for Ne, $3.80\times10^{-5}$ for Mg,
$3.55\times10^{-5}$ for Si, $1.62\times10^{-5}$ for S,
$3.63\times10^{-6}$ for Ar, $2.29\times10^{-6}$ for Ca,
$4.68\times10^{-5}$ for Fe, and $1.78\times10^{-6}$ for Ni (Anders,
Grevesse 1989).

%section 2
\section{Observation and Data Reduction}

ASCA observed M82 once in 1993, and observed it nine times
in 1996. Though the results of the observation in 1993 have
been reported by many authors (Tsuru et al. 1994; Awaki et
al. 1996; Tsuru et al. 1996; Moran, Lehnert 1997; Ptak et
al. 1997; Tsuru et al. 1997; Dahlem et al. 1998), we
reanalyzed the data in 1993 in the same method as for the
1996 data to investigate the time variability of M82
systematically. The dates of these observations are listed
in table 1.

All of the data were obtained with two solid-state imaging
spectrometers (SIS0 and SIS1) and two gas-imaging
spectrometers (GIS2 and GIS3) at the foci of four thin-foil
X-ray mirrors (XRT) on board the ASCA satellite. Details of
the instruments can be found in Burke et al. (1991), Ohashi
et al. (1996), Makishima et al. (1996), and Serlemitsos et
al. (1995), while Tanaka et al. (1994) gives a general
description of ASCA. The SIS data were obtained in the 4-CCD
bright mode for the observation in 1993 and in the 1-CCD
faint mode for the other observations in 1996. The GIS data
were obtained in the normal PH mode.

All of the data were screened with the standard selection
criteria: data taken in the South Atlantic Anomaly, Earth
occultation, and regions of low geomagnetic rigidity are
excluded. We also eliminated contamination by the bright
Earth, removed hot and flickering pixels from the SIS data,
and applied rise-time rejection to exclude particle events
from the GIS data. We further applied the ``flare-cut''
criteria to exclude the non X-ray background events as many
as possible for the GIS data (Ishisaki et al. 1997). After
these screenings, we obtained effective exposure times shown
in table 1.

%section 3
\section{Analysis and Results}

\subsection{Imaging Analysis}

Following Tsuru et al. (1997), we made the SIS images (SIS0
+ SIS1) in three energy bands (0.4 -- 0.8 keV, 1.2 -- 1.8
keV, and 3.0 -- 10 keV) for an individual observation. We
call them the soft-, medium-, and hard-band images,
respectively. The three energy bands represent the three
spectral components found by Tsuru et al. (1997). Though the
soft-band images show somewhat complex structures, we found
that all the medium- and hard-band images are single-peaked
and no other significant source exists in the field-of-view.

Tsuru et al. (1997) found from the PV data (April 19, 1993)
that the radial profiles of the soft- and medium-band images
are extended compared with the ASCA PSF, while the
hard-band image is consistent with a point source.  They
also found that the peak position of the hard-band image
agrees with that of the medium-band image, but does not
agree with that of the soft-band image.  We confirmed these
results of Tsuru et al. (1997) for all the ASCA data in
1996. For example, we show the SIS images of {\#74049000
(March 22, 1996)} in figure 1.  The soft-band peak is at
$\sim 0.\hspace{-2pt}'6$ southeast of the hard- and
medium-band peak.

The 90\% confidence level error circle of the pointing
position of ASCA is $\sim 1.\hspace{-2pt}'3$ diameter
(Gotthelf 1996). Except for {\#74049090} (November 26,
1996), the peak positions of hard-band images are consistent
among observations within $\sim 0.\hspace{-2pt}'8$. The
hard-band peak of {\#74049090} is located at $\sim
1.\hspace{-2pt}'4$ southeast of that of the PV data (April
19, 1993), which is still almost at the edge of the error
circle. Thus, we conclude that the hard-band peaks were
located at the same position within the error, and hence the
source at the hard-band peak is identical through all the
observations. The position of the hard-band peak determined
from the ASCA SIS data is $(\alpha, \delta)_{\rm J2000} =
(9^{\rm h}55^{\rm m}52^{\rm s}, 69^\circ40'48'')$.

\subsection{Light Curve}

From the screened data, we extracted the GIS and SIS light
curves for each observation from circular regions centered
on M82 in two energy bands (0.7 -- 1.5 keV and 3.0 -- 10.0
keV). The bin width of the light curves is 128 s. The
extraction radii are $6'$ for the GIS and $4'$ for the SIS,
which are nominal for a bright point source according to
``The ASCA DATA Reduction Guide'' presented by the ASCA
Guest Observer Facility\footnote{Please see
http://heasarc.gsfc.nasa.gov/docs/asca/abc/abc.html}. Next,
we combined the light curves of the SIS0 and SIS1, and the
GIS2 and GIS3.

We fitted the light curves of the GIS2 + GIS3 with a
constant counting rate model. In most cases, the reduced
$\chi^2\ (=\chi^2/d.o.f.)$ is less than 1.5, which means no
apparent time variability.  However, the reduced $\chi^2$ is
2.7 for the GIS light curve of {\#74049030 (April 24, 1996)}
in the 3.0 -- 10.0 keV band, though we can see no clear time
variability in the 0.7 -- 1.5 keV band at that time (figure
2). The SIS data are consistent with these results of the
GIS. As mentioned in the previous section, only one
identical source was found in the field-of-view through all
the observations. Therefore this result strongly suggests
that the hard component of M82 detected by Tsuru et al. 
(1997) has short-term variability on a time scale of $\sim\
(1 - 2)\times10^{4}$ s, while the soft component has no
clear time variability.

\subsection{Spectral Analysis}

From the screened data, we extracted the SIS and GIS spectra
from the same regions as those for the light curves.  Then
we rebinned the spectra to contain at least 20 counts in
each spectral bin to utilize the $\chi^2$ technique.

We extracted the background spectra for the GIS from the
blank sky data taken during the Large Sky Survey project
(e.g. Ueda et al. 1998) with the same data reduction method
for the data of M82. We obtained the SIS background spectra
for the PV data in 1993 from a source-free region around
M82. Because the observations in 1996 were conducted with
the 1-CCD faint mode and X-rays from M82 covered the whole
region of the CCD chip, we could not define the source-free
regions for the observations in 1996. Therefore, to extract
the background spectra, we used the data of LSS1988+317,
which is a dim source observed in the 1-CCD faint mode in
1996 (Sakano et al. 1998). After applying the same data
reduction as that for M82, we extracted the background
spectra from a source-free region.

The count rates of the spectra in the 0.7 -- 1.5 keV and 3.0
-- 10 keV bands are shown in figure 3.  They are background
subtracted but not vignetting corrected values.  The count
rates in the 3 -- 10 keV band show rather large time
variability compared with those in the 0.7 -- 1.5 keV band. 
This suggests that the hard component (Tsuru et al. 1997)
has long-term variability on a time scale of $\sim$ a month. 
Because the detector positions of M82 are almost the same
for all the observations in 1996, the vignetting correction
does not change the results essentially.

\subsubsection{Time variability of the hard component}

First, we fitted the SIS0 and SIS1 spectra in the 0.6 -- 10
keV band and the GIS2 and GIS3 spectra in the 0.7 -- 10 keV
band simultaneously with the three-temperature thermal
plasma model applied in Tsuru et al. (1997).  The model can
be represented by
%equation 1
\begin{equation}
N_{\rm H}(\mbox{whole}) \times [\mbox{RS(soft)}
+ N_{\rm H}(\mbox{medium}) \times \mbox{RS(medium)}
+ N_{\rm H}(\mbox{hard}) \times \mbox{RS(hard)}],
\end{equation}
where $N_{\rm H}$ is the equivalent hydrogen column density
of an absorbing cold material and RS is the thin thermal
plasma model developed by Raymond and Smith (1977)
(hereafter RS model). The soft and medium components are
assumed to have the same metal abundance, and the abundance
ratios among metals of the hard component are fixed to be
cosmic. Column densities, temperatures, and metal abundances
of the soft and medium components are fixed to the best-fit
values of Tsuru et al. (1997); $N_{\rm H}$(whole) =
$3.0\times10^{20}$ cm$^{-2}$, $N_{\rm H}$(medium) = 0.0
cm$^{-2}$, $kT$(soft) = 0.32 keV, $kT$(medium) = 0.95 keV,
He = C = 1.0 cosmic, N = 0.0 cosmic, O = 0.063 cosmic, Ne =
0.15 cosmic, Mg = 0.25 cosmic, Si = 0.40 cosmic, S = 0.47
cosmic, Ar = Ca = 0.0 cosmic, and Fe = Ni = 0.049 cosmic. 
Thus the free parameters are the column density,
temperature, metal abundance, normalization of the hard
component, and the normalizations of the soft and medium
components. This 3RS model could fit all the spectra quite
well. The best-fit parameters are shown in table 2. The
luminosities of the soft and medium components are quite
stable, while the hard component has clear time variability.
This is shown in figure 4.

The hard component could also be fitted with a power-law
model (Tsuru et al. 1997). Therefore, we also tried the 2RS
+ power-law model, which is expressed as
%equation 2
\begin{equation}
N_{\rm H}(\mbox{whole}) \times [\mbox{RS(soft)}
+ N_{\rm H}(\mbox{medium}) \times \mbox{RS(medium)}
+ N_{\rm H}(\mbox{hard}) \times \mbox{Power-law(hard)}].
\end{equation}
The soft and medium components are assumed to have the same
metal abundances. Column densities, temperatures and metal
abundance of the soft and medium components are fixed to the
best-fit values of Tsuru et al. (1997); $N_{\rm H}$(whole) =
$3.0\times10^{20}$ cm$^{-2}$, $N_{\rm H}$(medium) = $0.0$,
$kT$(soft) = 0.31 keV, $kT$(medium) = 0.95 keV, He = C = 1.0
cosmic, N = 0.0 cosmic, O = 0.061 cosmic, Ne = 0.14 cosmic,
Mg = 0.25 cosmic, Si = 0.40 cosmic, S = 0.45 cosmic, Ar = Ca
= 0.0 cosmic, and Fe = Ni = 0.048 cosmic. Thus the free
parameters were the column density, photon index,
normalization of the hard component, and the normalization
of the soft and medium components.  This 2RS + power-law
model also could fit all the GIS and SIS spectra. The
best-fit parameters are shown in table 3. Only the hard
component shows a time variability, which is the same as the
results of the 3RS model fitting.  Though both the 3RS and
2RS + power-law models could fit the spectra well, we should
note that the $\chi^2$ values of the 3RS model are generally
smaller than those of the 2RS + power-law model.

Figure 5 shows the GIS2 spectra of the highest state
(\#74049010; April 15, 1996) and the lowest state
(\#74049070; October 14, 1996). This figure also shows that
only the hard component has a significant time variability. 
Therefore, we can obtain the spectrum of the variable
component by subtracting the spectrum of the lowest state
from the highest state, which is shown in figure 6. The
spectrum could be fitted by either the heavily absorbed RS
or heavily absorbed power-law model (table 4).

\subsubsection{Iron K-line emission}

Information concerning the iron K-line emission is one of
the essential keys to determine the origin of the hard
component. Since the statistics of an individual observation
are limited, we added all the GIS spectra.  Since the PV
data in 1993 obtained in the 4-CCD bright mode in 1993 are
thought to be quite different in quality from those obtained
in the 1-CCD faint mode in 1996, we added only the data in
1996 for the SIS. Before adding the spectra, we confirmed
the accuracy of the gain calibration using the center energy
of the silicon K line at 1.86 keV. Then we simultaneously
analyzed the added SIS and GIS spectra and the SIS
spectra in 1993 above the 4 keV band.

First, we fitted the spectra with a thermal bremsstrahlung
model. The quality of the fit is good ($\chi^2/d.o.f.$ =
350.4/563), and is shown in figure 7 (a) and table 5. Next,
we added a Gaussian line model to the thermal bremsstrahlung
(Brems. + 1 Gaussian). The result is given in figures 7 (b),
8 and table 5, and the $\chi^2/d.o.f.$ is 330.8/560. The
decrease of the $\chi^2$ value is 19.6, while the decrease
of the $d.o.f.$ is 3. Therefore, we can conclude that ASCA
detected the iron K line with significant line width of
$\sigma \sim 0.3$ keV at a significance level of more than
99.5 \% (Malina et al. 1976).

Assuming the line broadening of 0.3 keV results from the
outflow motion of the hot gas, the velocity of the gas
should be $1.4\times10^9$cm/s, which is much too higher than
that estimated (e.g. McKeith et al. 1995). The detected line
center energy is between the iron 6.4 keV and 6.7 keV. Thus
the obtained broad line suggests a superposition of a few
lines, as seen in our galactic center (Koyama et al. 1996). 
Therefore, we added two narrow Gaussian lines ($\sigma$ = 0)
to the bremsstrahlung model (Brems. + 2 Gaussians). The
best-fit parameters are shown in table 5. In this case, the
two lines can be attributed to the iron 6.4 keV and 6.7 keV
lines. Finally, we tried three narrow Gaussian lines plus
the bremsstrahlung model in which the center energy is fixed
to 6.4 keV, 6.7 keV, and 7.0 keV (Brems. + 3 Gaussians). The
results are also shown in table 5.

%section 4
\section{Discussion}

\subsection{Identification of the Hard Component}

Two bright X-ray sources showing time variability have been
detected with the Einstein HRI and ROSAT HRI (Watson et al. 
1984; Collura et al. 1994); source 1 and 2 of Collura et al. 
(1994). We call them source 1 and source 2, hereafter. The
position of the hard component determined solely from the
ASCA SIS data is $(9^{\rm h}55^{\rm m}52^{\rm s},
69^\circ40'48'')_{\rm J2000}$ with an error circle of $\sim
40''$ radius in 90\% confidence level (Gotthelf 1996). The
error circle includes source 2. However, the position of the
hard component does not agree with source 1, whose position
is outside the error circle.

Next, we determine the precise position of the hard
component by comparing the ASCA SIS image with the ROSAT HRI
image.  Making the ASCA SIS image in the same energy range
as the ROSAT HRI and comparing it with the actual
ROSAT HRI image is insufficient, because the shape of
effective area as a function of X-ray energy is different
between the two instruments. Therefore, we make a fake ROSAT
HRI image from the ASCA SIS data by taking the energy
dependence of the effective area of the two instruments into
account, and comparing it with the actual ROSAT HRI image.

To fake the HRI image, we first made the four SIS images in
the 0.4 -- 0.8 keV, 0.8 -- 1.2 keV, 1.2 -- 1.6 keV, and 1.6
-- 2.0 keV bands. Next, we multiplied each image by the
ratio of the effective area of the HRI to that of the SIS in
each energy band. For the effective area of the ROSAT HRI,
we used the values in Briel et al. (1996).  For the SIS
image of the 0.4 -- 0.8 keV band, particularly, we multiply
it by the ratio of the effective area of the HRI in the 0.1
-- 0.8 keV band to that of the SIS in the 0.4 -- 0.8 keV
band, because the SIS has no efficiency below 0.4 keV. 
Finally, we made the fake HRI image by adding all four
images. In figure 9, we show the fake HRI image using the
data of {\#74049000 (March 22, 1996)} as an example. From
this analysis, we found that the peak position of the fake
HRI image agrees with the ASCA hard-band peak. Please note
that the fake HRI image almost reflects the ASCA medium-band
image because the peak of effective area as a function of
X-ray energy of the ROSAT HRI is at $\sim 1.1$ keV (Briel et
al. 1996). Because the peak of the fake HRI image should be
at the peak of the real HRI image, we conclude the ASCA
hard-band peak agrees with the peak of the real HRI image,
which is located at the nucleus of M82 (Collura et al. 1994;
Strickland et al. 1997).

Source 2 is time variable and located at the peak of the HRI
image. This source is reported to correspond to a strong 6 cm
radio source 41.5+597 (Strickland et al. 1997), which showed
a 100 \% drop in flux within a year (Muxlow et al. 1994). We
estimate the count rate of the ROSAT HRI of the hard
component from the best-fit parameters given in tables 2 and
3 and compare it with source 2. When we apply the RS model
for the hard component, the count rates of the ROSAT HRI are
estimated to be 0.017 c/s, 0.096 c/s, and 0.0040 c/s for the
PV data (April 19, 1993), the highest state (\#74049010;
April 15, 1996), and the lowest state (\#74049070; October
14, 1996), respectively. In the case of the power-law model,
the count rates are estimated to be 0.012 c/s for the PV
data, 0.075 c/s for the highest state, and 0.0040 c/s for
the lowest state. The range of these estimated counting
rates of the hard component includes the actual ROSAT HRI
counting rates of $0.04\sim 0.06$ c/s of source 2. Thus, we
conclude that the hard component at the ASCA hard-band peak
is identical with source 2.

\subsection{Origin of the Hard Component}

The hard component has time variability of $10^4$s -- a
month with luminosity of $3\times10^{40}$ --
$1\times10^{41}$ erg/s. A collection of discrete sources
such as X-ray binaries, a super Eddington source, a young
SNR, inverse Compton scattering of IR photons from
relativistic electrons which produce the radio emission, hot
interstellar medium, and an AGN have been considered so far
as an origin of the hard X-ray emission. Since the typical
X-ray luminosity of X-ray binaries is $\sim\ 10^{37}$ erg/s,
a few hundred binaries are needed to explain the luminosity
of the hard X-ray emission of M82.  Therefore, a collection
of binaries cannot explain the time variability of M82. 
Since the typical luminosity of super Eddington sources is
$\sim\ 10^{40}$ erg/s (Okada et al. 1998), it cannot explain
the highest luminosity of M82. A young SNR cannot become as
bright as M82 did in our observation. The radio emission is
extended as large as $\sim\ 30''$ (Kronberg, Clarke 1978),
which corresponds to $\sim\ 1.5\times10^{21}$ cm, while the
time variability on a time scale of $\sim\ 1\times10^{4}$ s
implies the size of the emitting region to be smaller than
$3\times10^{14}$ cm. Therefore the inverse Compton emission
cannot explain the time variability. It is also quite an
unnatural situation that the hot interstellar medium with
temperature of $\sim\ 10$ keV is confined to a region
smaller than $3\times10^{14}$ cm.  Therefore, an AGN is the
only possible origin that can explain the luminosity and
time variability.

The X-ray luminosity of the hard component is similar to
that of low-luminosity AGNs (LLAGN) (Terashima et al. 
1998b). We proved that the hard component is located at the
center of M82 and spatially unresolved. The residual
spectrum obtained by subtracting the spectrum of the lowest
state (\#74049070; October 14, 1996) from the highest state
(\#74049010; April 15, 1996) suggests the strong absorption,
which means that the variable source is embedded in the
galactic center region of M82. These also strongly support
the LLAGN origin of the hard component. Because the violent
starburst activity dominates the AGN activity, there has
been no evidence of the AGN in other wavelengths (e.g. Luts
et al. 1998), and one can detect the AGN activity only in
the hard X-ray band like NGC6240 (Iwasawa, Comastri 1998).
Though we could not constrain the center energy of the iron
K line, Cappi et al. (1998) determined the center energy to
be 6.7 keV, which may suggest the thermal origin of the hard
component. However, we should note that some of the LLAGNs
have the broad iron K line at 6.7 keV rather than at 6.4 keV
(Ishisaki et al. 1996; Terashima et al. 1998a).

As mentioned above, the time variability implies the size of
the emitting region to be smaller than $3\times10^{14}$ cm. 
Assuming that the origin of the hard component is the LLAGN
and the X-rays are mainly emitted from a region as large as
6 times its Schwarzschild radius, the mass of the central
object is estimated to be less than $2\times10^{8}\MO$. The
temperature, photon index, and column density showed in
tables 2 and 3 are also not constant, though there is no
clear correlation between them. However, there is a
correlation between the column density and luminosity of the
hard component; the column density increases as the
luminosity decreases (figure 10).  The same tendency can be
seen in NGC4151, which is the only known source showing the
variable column density (Yaqoob et al. 1993).

Though the most plausible origin of the hard component is a
LLAGN, the spectral shape is somewhat strange.  The $\chi^2$
values of the 3RS model fitting are generally smaller than
those of the 2RS + power-law model fitting.  The Ginga
spectrum could be fitted by the thermal bremsstrahlung model
but could not be fitted by the power-law model (Tsuru 1992). 
Almost the same conclusion was reported by Cappi et al. 
(1998) using the BeppoSAX data. All these suggest the
spectral shape is thermal-like, while typical LLAGN shows
power-law spectrum. Therefore, it may also be possible that
the origin of the hard component is a new type of accreting
X-ray sources.

%section 5
\section{Conclusion}

We found time variability of the hard component of M82
detected by Tsuru et al. (1997) at a time scale of $10^{4}$
s -- a month. The luminosity of the hard component in the
0.5 -- 10 keV band ranges from $3\times10^{40}$ erg/s to
$1\times10^{41}$ erg/s. The spatial position of the hard
component agrees with the luminous X-ray point source at the
center of M82 detected with ROSAT and Einstein (Watson et
al. 1984; Collura et al. 1994; Bregman et al. 1995;
Strickland et al. 1997). The spatial extent of the hard
component is consistent with a point source. The residual
spectrum obtained by subtracting the spectrum of the lowest
state from the highest state suggests the strong absorption
feature, which means that the variable source is embedded in
the galactic center region of M82. All these strongly
suggest that there is a hidden LLAGN in M82, though its
spectral shape is different from typical LLAGNs. We also
detected a broad iron K line at $6.56^{+0.14}_{-0.14}$ keV
or a superposition of a few iron K lines. There is a
correlation between the column density and luminosity of the
hard component.  This first firm evidence of a LLAGN in M82
suggests a link between starbursts and AGN.

\par

\vspace{1pc}\par

The authors thank K. Koyama, S. Ueno, Y. Terashima, and S. 
Yamauchi for helpful discussion and useful comments. They
are also grateful to P. Hilton for careful review of the
manuscript.  HM is supported by the Special Postdoctoral
Researchers Program of RIKEN.  The authors also thank the
ASCA team members for their support.

\vspace{1pc}\par
{\it Note added in proof.} -- 
After the submitting this paper, the authors have become aware of the
paper by Ptak and Griffiths (astro-ph/9903372), which obtains the
similar results to this paper using the same data independently.

%%%%%%%%%%%%%%%%

\clearpage
\section*{References}

\re
Anders E., Grevesse N. 1989, Geochim. Cosmochim. Acta 53, 197

\re
Awaki H, Tsuru T., Koyama K., {\it ASCA} team 1996, in UV
and X-Ray Spectroscopy of Astrophysical and Laboratory
Plasmas, ed. K. Yamashita, T. Watanabe (Universal
Academy Press, Tokyo), p327

\re
Bregman J. N., Schulman E., Tomisaka K. 1995, ApJ 439, 155

\re
Briel U. G., Aschenbach B., Hasinger G., Hippmann H.,
Pfeffermann E., Predehl P., Schmitt J. H. M. M., Voges W.
et al. 1996, The ROSAT User's Handbook

\re
Burke B. E., Mountain R. W., Harrison D. C., Bautz M. W.,
Doty J. P., Ricker G. R., Daniels P. J. 1991, IEEE Trans ED-38, 1069

\re
Cappi M., Persic M., Mariani S., Bassabu L., Danese L.,
Dean A. J., Di Cocco G., Franceschini A. et al. 1998, 
preprint (astro-ph/9809325)

\re
Collura A., Reale F., Schulman E., Bregman J. N. 1994,
ApJ 420, L63

\re
Dahlem M., Weaver K. A., Heckman T. M. 1998, ApJS 118, 401

\re
Gotthelf E. 1996, ASCA News No. 4, p31 (ASCA Guest Observer
Facility, NASA, Goddard Space Flight Center)

\re
Ishisaki Y., Makishima K., Iyomoto N., Hayashida K.,
Kohmura Y., Mushotzky R. F., Petre R., Serlemitsos P. J., 
Terashima Y. 1996, PASJ, 48, 237, 1996

\re
Ishisaki Y., Ueda Y., Kubo H., Ikebe Y., Makishima K.,
the GIS team 1997, ASCA News No. 5, p26 (ASCA Guest Observer
Facility, NASA, Goddard Space Flight Center)

\re
Iwasawa K., Comastri A. 1998, MNRAS 297, 1219

\re
Koyama K., Maeda Y., Sonobe T., Takeshima T., Tanaka Y.,
Yamauchi S. 1996, PASJ 48, 249

\re
Kronberg P. P., and Clarke J. N. 1978, ApJ 224, L51

\re 
Lutz D., Kunze D., Spoon H. W. W., Thornley M. D. 1998, A\&A
333, L75

\re
McKeith C. D., Greve A., Downes D., Prada F. 1995, A\&A 293,
703

\re
Makishima K., Tashiro M., Ebisawa K., Ezawa H., Fukazawa Y.,
Gunji S., Hirayama M., Idesawa E. et al. 1996, PASJ 48, 171

\re
Malina R., Lampton M., Bowyer S. 1976, ApJ 209, 678

\re
Moran E. C., Lehnert M. D. 1997, ApJ 478, 172

\re
Muxlow T. W. B., Pedlar A., Wilkinson P. N., Axon D. J.,
Sanders E. M., de Bruyn A. G. 1994, MNRAS 266, 455

\re
Ohashi T., Ebisawa K., Fukazawa Y., Hiyoshi K., Horii M., 
Ikebe Y., Ikeda H., Inoue H. et al. 1996, PASJ 48, 157

\re
Okada K., Dotani T., Makishima K., Mitsuda K., Mihara T. 1998,
PASJ 50, 25

\re
Ptak A., Serlemitsos P., Yaqoob T., Mushotzky R., Tsuru T.
1997, AJ 113, 1286

\re
Raymond J. C., Smith B. W. 1977, ApJS 35, 419

\re
Sakano M., Koyama K., Tsuru T., Awaki H., Ueda Y., Takahashi T.,
Akiyama M., Ohta K., Yamada T. 1998, ApJ 505, 129

\re
Schaaf R., Pietsch W., Biermann P. L., Kronberg P. P.,
Schmutzler T. 1989, ApJ 336, 722

\re
Serlemitsos P. J., Jalota L., Soong Y., Kunieda H., Tawara Y., 
Tsusaka Y., Suzuki H., Sakima Y. et al. 1995, PASJ 47, 105

\re
Strickland D. K., Ponman T. J., Stevens L. R. 1997, 
A\&A 320, 378

\re
Tanaka Y., Inoue H., Holt S. S. 1994, PASJ 46, L37

\re
Terashima Y., Kunieda H., Misaka K., Mushotzky R. F.,
Ptak A. F., Reichert G. A. 1998a, ApJ 503, 212

\re
Terashima Y., Kunieda H., Serlemitsos P. J., Ptak A. 1998b,
in THE HOT UNIVERSE (IAU symposium No. 188), ed. K. Koyama,
S. Kitamoto, M. Itoh (Kluwer Academic Publishers,
Dordrecht), p444

\re
Tsuru T. 1992, PhD Thesis, The University of Tokyo

\re
Tsuru T., Hayashi I., Awaki H., Koyama K., Fukazawa Y.,
Ishisaki Y., Iwasawa K., Ohashi T. et al. 1994, 
in New Horizon of X-Ray Astronomy,
ed. F. Makino, T. Ohashi (Universal Academy Press, Tokyo), p529

\re
Tsuru T., Hayashi I., Awaki H., Koyama K., Fukazawa Y.,
Ishisaki Y., Iwasawa K., Ohashi T. et al. 1996, 
in X-ray Imaging and Spectroscopy of Cosmic Hot Plasmas, 
ed. F. Makino, K. Mitsuda (Universal Academy Press, Tokyo), p157

\re
Tsuru T., Awaki H., Koyama K., Ptak A. 1997, PASJ 49, 619

\re
Ueda Y., Takahashi T., Inoue H., Tsuru T., Sakano M., Ishisaki Y.,
Ogasaka Y., Makishima K. et al. 1998, Nature 391, 866

\re
Watson M. G., Stanger V., Griffiths, R. E. 1984, ApJ 286, 144

\re
Yaqoob T., Warwick R. S., Makino F., Otani C., Sokoloski J. L.,
Bond A., Yamauchi M. 1993, MNRAS 262, 435

%%%%%%%%%%%%%%%%%%%%%%%

\clearpage

\centerline{Figure Caption}
\bigskip
\begin{fv}{1}{}
{SIS images of M82 from {\#74049000 (March 22, 1996)} without
background subtraction and vignetting correction, (a) in the 0.4 --
0.8 keV band, (b) in the 1.2 -- 1.8 keV band, and (c) in the 3.0 -- 10
keV band. The pixel size is $6.\hspace{-2pt}''4$, and the images have
been smoothed with a Gaussian distribution of $\sigma$ = 2 pixel. The
contour levels are 0.1, 0.2, 0.3, ...  c/pixel for (a), and 1, 2,
3,... c/pixel for both (b) and (c).  Dots, triangles, and stars denote
the peaks of the soft-, medium- and hard-band images, respectively.  }
\end{fv}

\begin{fv}{2}{}
{Light curves of the GIS2+3 during the observation of \#74049030
(April 24, 1996).  Each bin width is 128 s. The horizontal axis is
time after the start of the observation. No background subtraction is
applied.  The background levels (0.0079 c/s for the 0.7 -- 1.5 keV
band and 0.012 c/s for the 3.0 -- 10.0 keV band) are shown by the
dashed lines.}
\end{fv}

\begin{fv}{3}{}
{Count rates of the GIS2+3 in the 0.7 -- 1.5 keV and 3.0 -- 10 keV
bands.  The error bars are smaller than the data points.}
\end{fv}

\begin{fv}{4}{}
{Time variability of each component in the 3RS model. Stars, crosses,
and circles show the hard, medium, and soft components, respectively.}
\end{fv}

\begin{fv}{5}{}
{GIS2 spectra of the highest state (\#74049010; April 15,1996) and the
lowest state (\#74049070; October 14, 1996).}
\end{fv}

\begin{fv}{6}{}
{Residual spectra of the SIS0, SIS1, GIS2, and GIS3 obtained by
subtracting the lowest state (\#74049010; April 15, 1996) from the
highest state (\#74049070; October 14, 1996). The lines show the
best-fit heavily absorbed RS model.}
\end{fv}

\begin{fv}{7}{}
{The composite SIS and GIS spectra and the SIS spectra in 1993 above
the 4 keV band.  The solid lines show the best-fitting models: (a)
thermal bremsstrahlung model, and (b) thermal bremsstrahlung +
Gaussian line model.  }
\end{fv}

\begin{fv}{8}{}
{Confidence contours at $\Delta\chi^2$ = 2.3, 4.61, and 9.21 for the
line center energy and equivalent width of the iron K line.}
\end{fv}

\begin{fv}{9}{}
{The fake ROSAT HRI image made of the SIS images of {\#74049000 (March
22, 1996)} without background subtraction and vignetting correction. 
The pixel size is $6.\hspace{-2pt}''4$, and the image has been
smoothed with a Gaussian distribution of $\sigma$ = 2 pixel. The
contour levels are 1, 2, 3, ...  c/pixel.  The symbols are the same as
in Fig. 2.  }
\end{fv}

\begin{fv}{10}{}
{Column density of the hard component plotted against the luminosity
of the hard component in the 2 - 10 keV band. The column density and
luminosity are evaluated with the 3RS model.}
\end{fv}

%\end{document}

%%%%%%%%%%%%%%%%%%%%%%%
\clearpage

%table 1
\begin{table}[t]

\begin{center}
Table~1.\hspace{4pt}Log of the observations.
\end{center}

\begin{tabular*}{\textwidth}{@{\hspace{\tabcolsep}\extracolsep{\fill}}p{6pc}ccccccc} \hline \hline
Sequence	&\multicolumn{2}{c}{Date}	&	&\multicolumn{4}{c}{Exposure}     \\
\cline{2-3}	\cline{5-8}
		&UT	&MJD			&	&SIS0	&SIS1	&GIS2	&GIS3\\
		&(dd/mm/yy)	&		&	&(s)	&(s)	&(s)	&(s)\\ \hline

PV \dotfill	&19/04/1993	&49096.5	&	&16254	&16521	&26694	&26646	\\

74049000 \dotfill	&22/03/1996	&50164.5	&	&13565	&13565	&12932	&12924	\\

74049010 \dotfill	&15/04/1996	&50188.5	&	&6930	&6930	&8668	&8668	\\

74049020 \dotfill	&21/04/1996	&50194.5	&	&12395	&12343	&13140	&13138	\\

74049030 \dotfill	&24/04/1996	&50197.5	&	&29800	&29902	&33294	&33288	\\

74049050 \dotfill	&13/05/1996	&50216.5	&	&8160	&8088	&9688	&9688	\\

74049060 \dotfill	&16/05/1996	&50219.5	&	&11545	&11583	&14386	&14390	\\

74049070 \dotfill	&14/10/1996	&50370.5	&	&8941	&8941	&7612	&7612	\\

74049080 \dotfill	&14/11/1996	&50401.5	&	&12033	&12103	&12498	&12496	\\

74049090 \dotfill	&26/11/1996	&50413.5	&	&8762	&8730	&9792	&9760	\\ \hline
\end{tabular*}

\end{table}

%%%%%%%%%%%%%%%%%%%%%%%%%
\clearpage

%table 2
\begin{table}[t]

\begin{center}
Table~2.\hspace{4pt}Results of the 3RS model fitting.
\end{center}

\scriptsize

\begin{tabular*}{1.1\textwidth}{@{\hspace{\tabcolsep}\extracolsep{\fill}}p{4pc}ccccccccccc} \hline \hline
Sequence	&$N_{\rm H}$	&kT	&Abundance	
&\multicolumn{3}{c}{$F_X^*$($10^{-11}$ erg/s/cm$^2$)}	&	
&\multicolumn{3}{c}{$L_X^{\dag}$($10^{40}$ erg/s)}	&
$\chi^2/d.o.f.$\\
\cline{5-7}\cline{9-11}
	&$10^{22}$ cm$^{-2}$	&keV	&cosmic
&soft	&medium	&hard$^{\S}$	&	&soft	&medium	&hard$^{\S}$\\ \hline

PV \dotfill		
&$2.20^{+0.34}_{-0.33}$	&$11.2^{+2.0}_{-1.6}$	&$0.18^{+0.08}_{-0.06}$
&0.44	&0.70	&2.0 (1.9)	&	&1.8	&1.6	&4.2 (2.9)	&1053.6/1022	\\

74049000 \dotfill			
&$1.89^{+0.31}_{-0.29}$	&$4.47^{+0.41}_{-0.38}$	&$0.036^{+0.042}_{-0.036}$
&0.34	&0.88	&2.2 (2.1)	&	&1.4	&2.0	&5.6 (3.2)	&796.2/840	\\

74049010 \dotfill			
&$1.35^{+0.25}_{-0.23}$	&$10.2^{+1.5}_{-1.3}$	&$0.026^{+0.071}_{-0.026}$
&0.41	&0.76	&5.5 (5.2)	&	&1.6	&1.7	&11 (7.4)	&859.1/870	\\

74049020 \dotfill			
&$1.38^{+0.19}_{-0.18}$	&$8.17^{+0.81}_{-0.73}$	&$0.033^{+0.044}_{-0.033}$
&0.43	&0.73	&4.5 (4.2)	&	&1.7	&1.7	&9.1 (6.1)	&1097.7/1065	\\

74049030 \dotfill			
&$1.47^{+0.13}_{-0.13}$	&$7.40^{+0.41}_{-0.38}$	&$0.036^{+0.026}_{-0.027}$
&0.42	&0.74	&4.4 (4.1)	&	&1.7	&1.7	&9.2 (6.0)	&1739.1/1594	\\

74049050 \dotfill			
&$1.74^{+0.56}_{-0.54}$	&$16.0^{+7.0}_{-4.6}$	&$0.00^{+0.14}_{\mbox{$\cdot \cdot \cdot$}}$
&0.34	&0.82	&2.3 (2.2)	&	&1.4	&1.9	&4.4 (3.2)	&634.0/596	\\

74049060 \dotfill			
&$1.58^{+0.30}_{-0.28}$	&$8.08^{+1.14}_{-0.90}$	&$0.075^{+0.060}_{-0.063}$
&0.34	&0.85	&2.8 (2.6)	&	&1.4	&1.9	&5.8 (3.9)	&929.9/889	\\

74049070 \dotfill			
&$3.66^{+0.99}_{-0.91}$	&$5.70^{+2.20}_{-1.32}$	&$0.22^{+0.13}_{-0.12}$
&0.33	&0.82	&1.2 (1.1)	&	&1.3	&1.9	&3.1 (2.0)	&338.1/315	\\

74049080 \dotfill			
&$2.84^{+0.73}_{-0.72}$	&$11.7^{+6.8}_{-3.2}$	&$0.12^{+0.13}_{-0.12}$
&0.36	&0.77	&1.7 (1.7)	&	&1.5	&1.8	&3.7 (2.6)	&544.7/479	\\

74049090 \dotfill			
&$1.76^{+0.75}_{-0.79}$	&$11.9^{+5.7}_{-3.1}$	&$0.090^{+0.13}_{-0.090}$
&0.37	&0.80	&2.6 (2.5)	&	&1.5	&1.8	&5.1 (3.6)	&490.9/466	\\ \hline

\end{tabular*}

\noindent
All the errors are described at 90\% confidence limits.

\noindent
$*$ Flux in the 0.5 -- 10 keV band.

\noindent
$\dag$ Unabsorbed luminosity in the 0.5 -- 10 keV band.

\noindent
$\S$ Values in parentheses are in the 2 -- 10 keV band.
\end{table}

%%%%%%%%%%%%%%%%%%%%%%%%%%%%%%%%%%%
%\clearpage

%table 3
\begin{table}[t]
\begin{center}
Table~3.\hspace{4pt}Results of the 2RS + power-law model fitting.
\end{center}

\scriptsize

\begin{tabular*}{1.05\textwidth}{@{\hspace{\tabcolsep}\extracolsep{\fill}}p{4pc}cccccccccc} \hline \hline
Sequence	&$N_{\rm H}$	&Photon Index
&\multicolumn{3}{c}{$F_X^*$($10^{-11}$ erg/s/cm$^2$)}
&
&\multicolumn{3}{c}{$L_X^{\dag}$($10^{40}$ erg/s)}
&$\chi^2/d.o.f.$\\
\cline{4-6}\cline{8-10}
	&$10^{22}$ cm$^{-2}$	&
&soft	&medium	&hard$^{\S}$	&	&soft	&medium	&hard$^{\S}$ \\ \hline

PV \dotfill				
&$2.87^{+0.41}_{-0.40}$	&$1.79^{+0.08}_{-0.08}$	
&0.42	&0.75	&2.0 (1.9)	&	&1.7	&1.7	&5.1 (3.1)	&1080.4/1023	\\

74049000 \dotfill			
&$3.07^{+0.40}_{-0.37}$	&$2.51^{+0.11}_{-0.10}$	
&0.30	&0.98	&2.2 (2.1)	&	&1.3	&2.2	&11 (3.7)	&817.2/841	\\

74049010 \dotfill			
&$1.90^{+0.31}_{-0.29}$	&$1.83^{+0.08}_{-0.07}$	
&0.37	&0.88	&5.5 (5.2)	&	&1.5	&2.0	&13 (7.9)	&876.9/871	\\

74049020 \dotfill			
&$2.00^{+0.24}_{-0.23}$	&$1.96^{+0.06}_{-0.06}$	
&0.40	&0.84	&4.5 (4.2)	&	&1.6	&1.9	&12 (6.5)	&1121.2/1066	\\

74049030 \dotfill			
&$2.25^{+0.17}_{-0.17}$	&$2.04^{+0.04}_{-0.04}$	
&0.37	&0.89	&4.4 (4.1)	&	&1.5	&2.0	&12 (6.5)	&1832.8/1595	\\

74049050 \dotfill			
&$2.35^{+0.70}_{-0.64}$	&$1.69^{+0.14}_{-0.13}$	
&0.32	&0.87	&2.3 (2.2)	&	&1.3	&2.0	&5.1 (3.3)	&637.1/597	\\

74049060 \dotfill			
&$2.23^{+0.38}_{-0.34}$	&$1.96^{+0.08}_{-0.09}$	
&0.31	&0.92	&2.8 (2.7)	&	&1.3	&2.1	&7.5 (4.1)	&949.6/890	\\

74049070 \dotfill			
&$4.27^{+1.17}_{-1.06}$	&$2.10^{+0.26}_{-0.24}$	
&0.32	&0.83	&1.2 (1.2)	&	&1.3	&1.9	&4.3 (2.1)	&347.5/316	\\

74049080 \dotfill			
&$3.65^{+0.85}_{-0.83}$	&$1.80^{+0.18}_{-0.16}$	
&0.34	&0.81	&1.7 (1.7)	&	&1.4	&1.8	&4.6 (2.8)	&549.5/480	\\

74049090 \dotfill			
&$2.75^{+0.77}_{-0.79}$	&$1.82^{+0.16}_{-0.15}$	
&0.32	&0.90	&2.6 (2.5)	&	&1.3	&2.1	&6.5 (3.9)	&501.2/467	\\ \hline

\end{tabular*}

\noindent
All the errors are described at 90\% confidence limits.

\noindent
$*$ Flux in the 0.5 -- 10 keV band.

\noindent
$\dag$ Unabsorbed luminosity in the 0.5 -- 10 keV band.

\noindent
$\S$ Values in parentheses are in the 2 -- 10 keV band.
\end{table}

%%%%%%%%%%%%%%%%%%%%%%%%%%%

\clearpage

%table 4
\begin{table}[t]

\begin{center}
Table~4.\hspace{4pt}The results of the fitting of the residual spectra.
\end{center}

\scriptsize

\begin{tabular*}{\textwidth}{@{\hspace{\tabcolsep}\extracolsep{\fill}}p{10cm}cc} \hline \hline
			&RS 		&power-law \\ \hline
$N_{\rm H}$ ($10^{22}$ cm$^{-2}$)
			&$1.30^{+0.12}_{-0.10}$	&$1.54^{+0.16}_{-0.14}$\\
Temperature (keV) \dotfill	&$11.7^{+2.6}_{-2.1}$	&$\cdot\cdot\cdot$\\
Abundance (cosmic) \dotfill	&$0.00^{+0.088}_{\cdot\cdot\cdot}$	&$\cdot\cdot\cdot$\\
Photon Index \dotfill	&$\cdot\cdot\cdot$	&$1.72^{+0.09}_{-0.08}$\\
Flux (0.5--10keV) (erg/s/cm$^2$) \dotfill	
			&$4.5\times10^{-11}$	&$4.6\times10^{-11}$\\
Flux (2--10keV) (erg/s/cm$^2$)	\dotfill
			&$4.2\times10^{-11}$	&$4.3\times10^{-11}$\\
Luminosity (0.5--10keV) (erg/s)	\dotfill
			&$8.3\times10^{40}$	&$9.5\times10^{-11}$\\
Luminosity (2--10keV) (erg/s) \dotfill
			&$5.8\times10^{40}$	&$6.1\times10^{-11}$\\
$\chi^2/d.o.f.$ \dotfill	&272.0/274	&279.5/275\\ \hline
\end{tabular*}

\noindent
All the errors are described at 90\% confidence limits.

\end{table}

%%%%%%%%%%%%%%%%%%%%%%%%%%%%

\clearpage

%table 5
\begin{table}[t]

\scriptsize

\begin{center}
Table~5.\hspace{4pt}The best-fit parameters of the model fitting
to the spectra above the 4 keV band.
\end{center}

\begin{tabular*}{1.1\textwidth}{@{\hspace{\tabcolsep}\extracolsep{\fill}}p{8pc}ccccccc} \hline\hline
	&Brems.	&Brems. + 1 Gaussian	&\multicolumn{2}{c}{Brems.+ 2 Gaussians}
&\multicolumn{3}{c}{Brems. + 3 Gaussians}\\ \hline

Temperature \dotfill		
&$12.56^{+1.95}_{-1.33}$	&$10.19^{+1.84}_{-1.05}$	
&\multicolumn{2}{c}{$10.71^{+1.43}_{-1.14}$}
&\multicolumn{3}{c}{$10.51^{+1.48}_{-1.16}$}\\

Center Energy (keV) \dotfill	
&\mbox{$\cdot\cdot\cdot$}	&$6.56^{+0.14}_{-0.14}$	
&$6.34^{+0.12}_{-0.14}$	&$6.68^{+0.17}_{-0.11}$
&$6.4^{\dag}$	&$6.7^{\dag}$	&$7.0^{\dag}$ \\

Line Width ($\sigma$) (keV) \dotfill	
&\mbox{$\cdot\cdot\cdot$}	&$0.30^{+0.16}_{-0.19}$
&$0^{\dag}$	&$0^{\dag}$
&$0^{\dag}$	&$0^{\dag}$	&$0^{\dag}$ \\

Equivalent Width (eV) \dotfill	
&\mbox{$\cdot\cdot\cdot$}	&$121^{+58}_{-61}$
&$39^{+26}_{-25}$	&$52^{+31}_{-30}$
&$45^{+26}_{-25}$	&$39^{+31}_{-31}$	&$22^{+37}_{-22}$\\

$\chi^2/d.o.f.$ \dotfill		
&350.4/563	&330.8/560	&\multicolumn{2}{c}{330.2/559}	
&\multicolumn{3}{c}{330.0/560}\\ \hline

\end{tabular*}

\noindent
All the errors are described at 90\% confidence limits.

\noindent
$\dag$ Fixed.

\end{table}

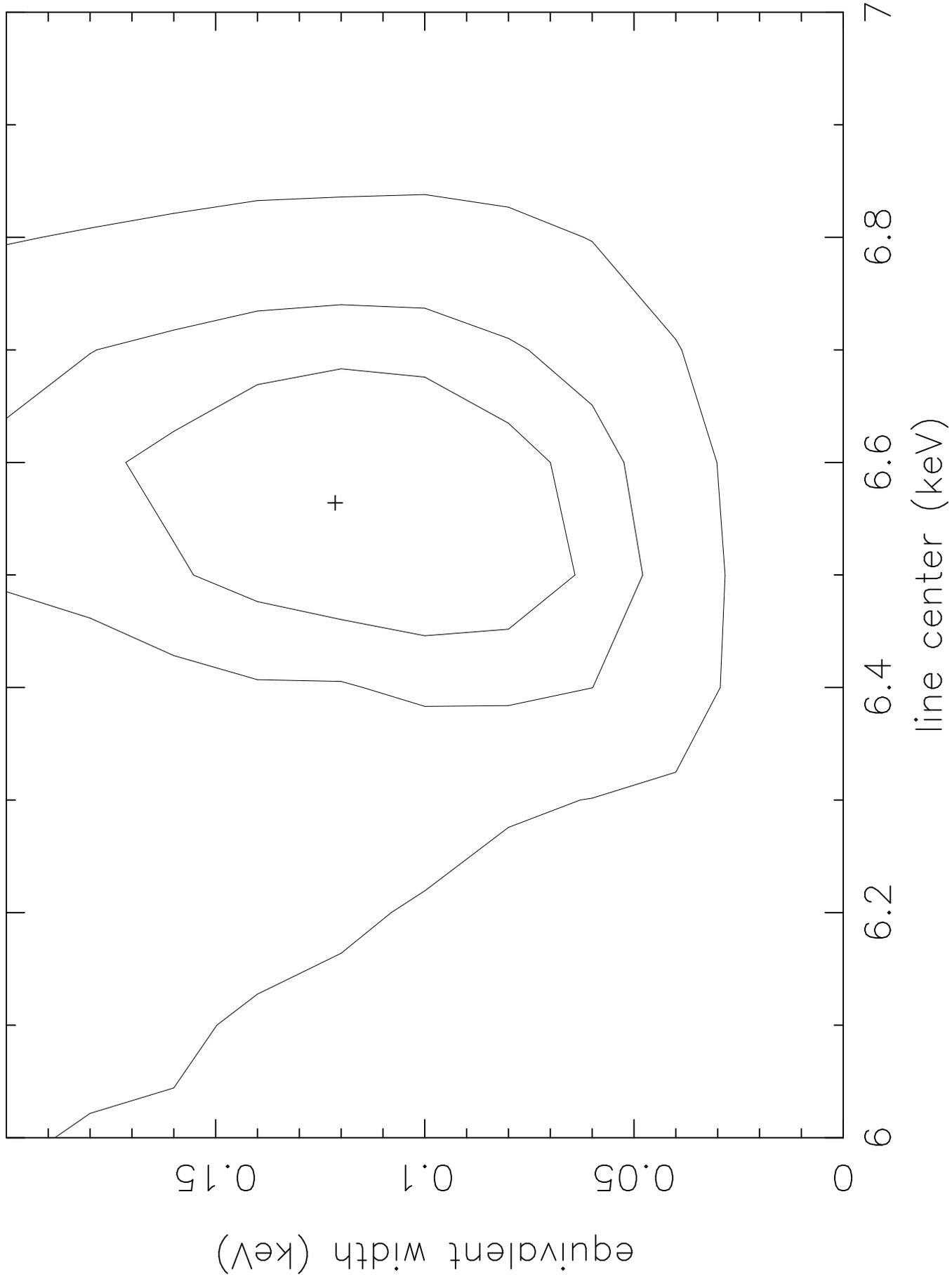
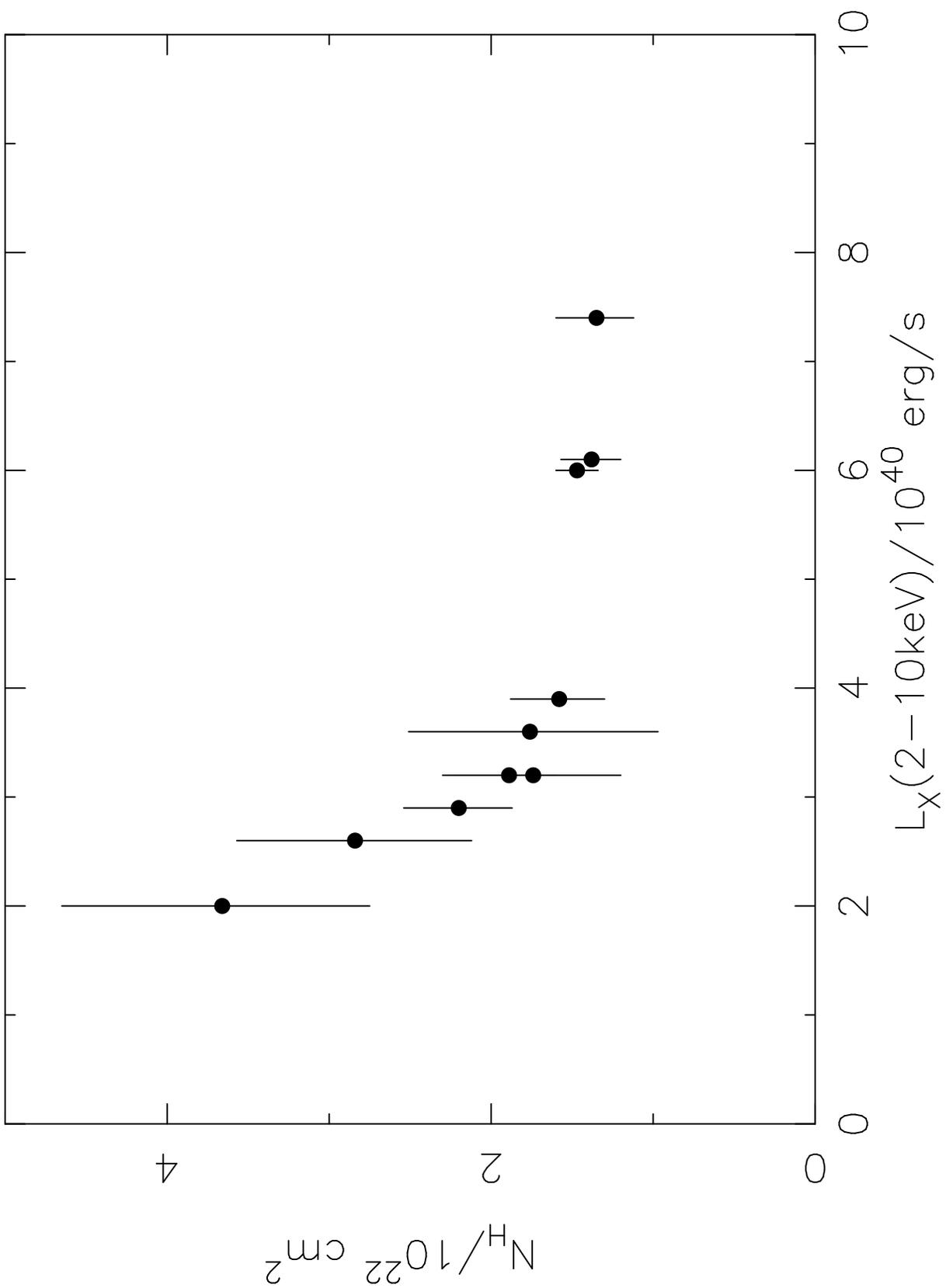
\end{document}